\documentclass[12pt]{article}
\usepackage{ulem}
\usepackage{graphicx,float,amsmath,amssymb,mathrsfs}
\usepackage{url,hyperref}
\usepackage[usenames,dvipsnames]{color}

 \usepackage{ulem}
\usepackage{graphicx,float,amsmath,amssymb,mathrsfs}
\usepackage[usenames,dvipsnames]{color}

\newcommand{\ket}[1]{|\kern.3ex#1\kern.3ex\rangle}
\newcommand{\bra}[1]{\langle\kern.3ex #1 \kern.3ex|}
\newcommand{\scalar}[2]{\langle\kern.3ex{#1}\kern.3ex|\kern.3ex{#2}\kern.3ex\rangle}

\newcommand \be  {\begin{equation}}
\newcommand \bea {\begin{eqnarray} \nonumber }
\newcommand \ee  {\end{equation}}
\newcommand \eea {\end{eqnarray}}

\renewcommand{\Re}{\mathop{\rm Re}}
\renewcommand{\Im}{\mathop{\rm Im}}

\begin{document}

\title{On the density of complex eigenvalues of Wigner reaction matrix in a disordered or chaotic system with absorption}

\vskip 0.2cm

\author{Yan V. Fyodorov $^1$ \\
\noindent\small{$^1$ King's College London, Department of Mathematics, London  WC2R 2LS, United Kingdom}}

\maketitle

\begin{abstract}
In an absorptive system the Wigner reaction $K-$matrix (directly related to the impedance matrix in acoustic or electromagnetic wave scattering) is non-selfadjoint, hence its
eigenvalues are complex. The most interesting regime arises when the absorption, taken into account as an imaginary part of the spectral parameter, is of the order of the mean level spacing. I show how to derive the mean density of the complex eigenvalues for reflection problems in disordered or chaotic systems with broken time-reversal invariance. The computations are done in the framework of nonlinear $\sigma-$ model approach, assuming fixed $M$ and $N\to \infty$.  Some explicit formulas are provided for zero-dimensional quantum chaotic system as well as for a semi-infinite quasi-1D system with fully operative Anderson localization.

\end{abstract}

\section{\small Introduction}
 Consider the problem of wave scattering from a piece of
 random medium confined to a spatial domain ${\cal D}$ and described by a self-adjoint Hamiltonian $H$, e.g.
\be\label{tightbinding}
H=\sum_{{\bf r}\in \mathbf{\Lambda}}\,V({\bf r})\left|{\bf r}\right\rangle\left\langle{\bf r}\right|+\sum_{{\bf r}\sim {\bf r}'}\left(t_{{\bf r}{\bf r}'}\left.|{\bf r}\right\rangle\left\langle {\bf r}'\right|+t_{{\bf r}'{\bf r}}\left|{\bf r}'\right\rangle\left\langle {\bf r}\right|\right),
\ee
where the second sum runs over nearest neighbours on a lattice $\mathbf{\Lambda}$ (assumed to be confined to the domain ${\cal D}$).  The parameters $t_{{\bf r}'{\bf r}}$ are in general complex satisfying $t^*_{{\bf r}{\bf r}'}=t_{{\bf r}'{\bf r}}$ to ensure the Hermiticity of the Hamiltonian: $H=H^{\dagger}$, where we use $t^*$ to denote complex conjugation of $t$ and $H^{\dagger}$ for Hermitian conjugation of $H$.
The disordered nature of the medium is taken into account by choosing the on-site potentials $V({\bf r})$ and/or hopping parameters $t_{{\bf r}'{\bf r}}$ to be random variables. Such construction is known in the literature as the Anderson model, and provides the paradigmatic framework to study single-particle localization phenomena. Note that the form
\eqref{tightbinding} can be used also for modelling a quantum particle
motion on any graph ${\bf r}\in \mathfrak{G}$ , with $t_{{\bf r}{\bf r}'}$ being the elements of the adjacency matrix of the graph $\mathfrak{G}$

The tight-binding representation is convenient as it allows one to think of such a Hamiltonian as described by a large $N\times N$ random matrix $H$, with $N$ being the number of sites in the lattice or graph. Alternatively, one may think of its continuum analogue, $H=-\frac{\hbar^2}{2m}\nabla^2+V({\bf r}), \quad {\bf r}\in {\cal D}$,
 with the appropriate (e.g. Dirichlet) conditions at the boundary of ${\cal D}$. In fact, under appropriate conditions the essentially random nature of wave scattering can be
 generated by an irregularly shaped boundary of the domain ${\cal D}$, without any intrinsic potential disorder.
   This is the standard case in the so-called wave billiards, the paradigmatic toy systems to study effects of quantum or wave chaos, see e.g. \cite{kuhl13,grad14,diet15,hcao15}. In such a case the famous Bohigas-Giannoni-Schmidt conjecture \cite{BGS1984} allows to describe universal features of such systems efficiently by replacing the Hamiltonian $H$ with a random $N\times N$ matrix from Gaussian
 Ensembles:  Gaussian Orthogonal (GOE),  Gaussian Symplectic (GSE), or Gaussian Unitary (GUE), depending on the presence or absence of time-reversal symmetry (and/or other relevant symmetries) in the system.

     A very convenient framework for describing scattering of classical or quantum waves from the disordered or chaotic medium has been formulated in Ref.\ \cite{VWZ85}, see e.g.\ \cite{FyoSom97} for more detail.
 Within such a framework, which is frequently called in the literature the ``Heidelberg model'',  one constructs  the unitary  $M\times M$ energy-dependent scattering matrix $S(E)$  describing scattering of waves incident on  a random medium at some energy $E$ and then exiting it via $M$ open scattering channels, numbered by $c=1,\ldots,M$, see a sketch below. Unitarity reflects the flux conservation: the vectors ${\bf a}=(a_1,\ldots,a_M)$ of incoming and ${\bf b}=(b_1,\ldots,b_M)$ of outgoing
amplitudes are linearly related via ${\bf b}=S(E)\,{\bf a}$ and have the same norm.

\begin{figure}
\centering
\includegraphics[width=70mm]{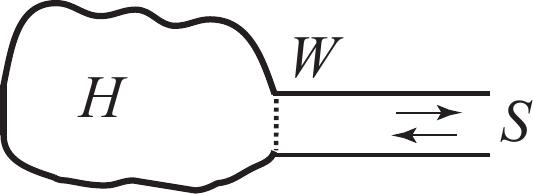}.
\caption{A sketch of a chaotic wave scattering from a region schematically represented by a cavity and assumed to contain a random medium inside.
 An operator governing wave dynamics in such a cavity
decoupled from the channels is assumed to be effectively described by a large random matrix $H$. An infinite lead is assumed to support $M$ propagating channels
in the considered energy range, and is coupled to the cavity region by a matrix/operator $W$. The  ensuing $M\times M$ unitary scattering matrix $S$
can be related to $H$ and $W$ in the framework of the Heidelberg approach, and is given by Eq.\ (\ref{1}).}
\label{F:sketch}
\end{figure}

 The relation between  $S(E)$ and the medium Hamiltonian $H$  is then provided by the following expression
\begin{equation}\label{1}
S(E)=\frac{{\bf 1}-i\mathbf{K}}{{\bf 1}+i\mathbf{K}}, \qquad
\mathbf{K}(E)=W^{\dagger}\frac{1}{E-H}W
\end{equation}
where columns $W_c$ ($c=1,\ldots, M$) of an $N\times M$ matrix $W$ of  {\it coupling amplitudes}\/ to $M$ open scattering channels can be taken as fixed vectors satisfying
the orthogonality condition:
\be \label{orth}
\sum_{i=1}^N W^*_{ci}W_{bi}=\gamma_c\delta_{cb},
\ee
 with $\gamma_c>0$ $\forall c=1,\ldots, M $  determining the ``bare'' strength of coupling of a given channel to the scattering system. The resulting $M\times M$  Hermitian matrix $\mathbf{K}(E)$ is known in the literature as the Wigner reaction $K-$ matrix.
 It is Hermiticity of $\mathbf{K}$ which implies $S-$matrix unitarity, hence implies the flux conservation. Note that the Wigner $K-$matrix is experimentally measurable in microwave scattering systems, as it is directly related to the systems impedance matrix, see e.g. \cite{Hemmadi2005,Hemmadi2006a,Hemmadi2006b}.

 One of serious challenges related to theoretical description of scattering characteristics is however related
 to the fact that experimentally measured quantities suffer from inevitable energy losses (absorption), e.g. due to
 damping in resonator walls and other imperfections. Such losses violate unitarity of the scattering matrix and are important for
 interpretation of experiments, hence considerable efforts were directed towards incorporating them into the Heidelberg approach
 \cite{Fyo05}. At the level of the model (\ref{1}) the losses can be taken into account by allowing the spectral parameter $E$ to have finite imaginary part by replacing $E\rightarrow E+i\alpha\in\mathbb{C}$ with some $\alpha>0$.
 This replacement violates Hermiticity of the  Wigner matrix $\mathbf{K}(E+i\alpha)$; in particular entries of $\mathbf{K}$ become now complex even for real symmetric choice of $H$ and real $W$. The most interesting, difficult and experimentally relevant regime occurs when absorption parameter $\alpha$ is comparable with the mean separation $\Delta(E)$ between neighbouring eigenvalues of the wave-chaotic Hamiltonian $H$. For example, if one uses the Gaussian random matrix model for $H$, normalized to have the mean eigenvalue density given by Wigner semicircle $\nu(E)=1/(2\pi)\sqrt{4-E^2}$ in a finite interval $|E|<2$,  one has  $\Delta(E)=(\nu(E)N)^{-1}$ as $N\to \infty$.  The statistics of the real and imaginary parts of $K-$matrix entries in such a regime was the subject of a considerable number of
 theoretical papers \cite{Fyo2003,FyoSav2004,SavSomFyo2005,SavFyoSom2006,FFWigner}  and by now well-understood  and measured experimentally with good precision  for systems with preserved time-reversal invariance in microwave cavities \cite{Kuhletal2003,Hemmadi2005,Hemmadi2006a,Hemmadi2006b} and microwave simulations of quantum graphs \cite{microwgraphs1,microwgraphs1a,microwgraphs2,microwgraphs3}. More recently experimental results for K-matrices in systems with broken time-reversal invariance  \cite{LawSir2018,LawSir2020} and eventually symplectic symmetry \cite{LawSymplec2023} have been also reported.

 In the present paper we will be interested in yet another characteristics of the non-Hermitian Wigner matrix $\mathbf{K}(E+i\alpha)$, the mean density of its complex eigenvalues
  $K_c=\Re K_c-i\Im K_c$, $\forall c=1,\ldots, M$ defined as
\begin{equation}\label{denK}
 \rho_M(u,v;y)=\left\langle \sum_{c=1}^M\delta\left(u-\Re K_c\right) \delta\left(v-\Im K_a\right)\right\rangle,
\end{equation}
where we suppressed the energy dependence for simplicity, indicating instead explicit dependence on the appropriately scaled absorption parameter $y=\frac{2\pi \alpha}{\Delta}$.
Here and henceforth the angular brackets $\left\langle \ldots \right \rangle$ indicate the averaging over ensemble of random Hamiltonians $H$. Note that by selecting
 the coupling vectors $W_c$ coinciding with the first $M$ basis vectors in $N-$dimensional space, i.e. $W_1={\bf e}_1=(1,0,\ldots,0),\,W_2={\bf e}_2=(0,1,0,\ldots,0)$, etc. converts the $K$-matrix to $M\times M$ top left corner of the $N\times N$  resolvent matrix  $(E+i\alpha -H)^{-1}$. Physically, this corresponds to $M$ perfectly coupled channels, attached to first $M$ sites.  From that angle we aim to characterize the eigenvalue density for
 the corner resolvent minor at complex values of the spectral parameter, an interesting and potentially rich mathematical problem. We are not aware of any systematic studies
 in that direction.

 Note that in fully chaotic, zero-dimensional system the positions of channel attachement do not play any role due to inherent ergodicity.
 In a more general non-ergodic situation, which may arise due to presence of Anderson localization phenomena, one may think of such an arrangement as corresponding to a wave reflection problem. In such setting the density  Eq.(\ref{denK}) has appeared recently in the paper \cite{FSTresden} as an important quantity facilitating computation of  the mean density of S-matrix poles, also known as {\it resonances}, in the complex energy plane. The latter density is experimentally measurable in wave-chaotic system \cite{kuhl_resonance_2008,CAF_2021_B} and
is a subject of long-standing theoretical interest, see e.g. \cite{SokZel89,Haake1992,Fyo96,FyoKhor99,Som99,scho00,FM2002}.  Clearly, the density  Eq.(\ref{denK})
is also experimentally measurable in principle, provided accurate experimental data can be sampled for the whole $K-$matrix.
  The paper \cite{FSTresden} included, without a proper derivation, an explicit expression for such density, valid for a general class disordered systems
  with broken time-reversal invariance, namely for those which can be mapped on the so-called supersymmetric nonlinear $\sigma-$model, see \cite{Efe_book}
  and discussions in  \cite{FSTresden} for more information. The present paper aims to fill in that gap by providing a detailed derivation, which involves several steps
   only relatively sketchy described in the available literature. 

To begin with, for the special simplest case $M=1$ the $K-$matrix consists of a single element, and finding the density Eq.(\ref{denK}) is equivalent to computing the
joint probability density of the real and complex parts of such an element. Such density has been originally addressed in \cite{MirFyo1994B} via quite a tediuos calculations in the $\sigma-$model approximation. A much more efficient approach has been proposed later in  \cite{SavSomFyo2005}, see also an account in \cite{Fyo05}. Our goal in this paper is to show how to generalize that approach to any number of open channels $M$, on an example of systems with broken time-reversal invariance.  Along these lines we also try to elucidate some features of the method which were omitted in the exposition of \cite{SavSomFyo2005,Fyo05}.

\section{\small Derivation of the main results}

\subsection{\small General exposition of the method}

Given two real parameters $p\in \mathbb{R}$ and $q>0$ we start with defining the following object:
 \begin{equation}\label{corr1}
 C_{\alpha}(p,q):=\left\langle Tr\left(z-\mathbf{K}(E+i\alpha)\right)^{-1}Tr\left(\overline{z}-\mathbf{K}(E-i\alpha)\right)^{-1}\right\rangle,
 \end{equation}
where we denoted $ z=p+iq, \, \overline{z}=p-iq$ and assumed the real energy $E$ and the absorption parameter $\alpha>0$ to be fixed.
As eigenvalues of the matrices  $\mathbf{K}(E+i\alpha)$ and $\mathbf{K}(E-i\alpha)$ are complex conjugates of each other, one
can write each trace in terms of $K_c=\Re K_c-i\Im K_c:=u_c-iv_c$, with $v_c>0$, representing Eq.(\ref{corr1}) as a sum of diagonal and off-diagonal contributions:
\[
 C_{\alpha}(p,q)= C^{(diag)}_{\alpha}(p,q)+C^{(off)}_{\alpha}(p,q)
\]
where
 \begin{equation}\label{corr2diag}
 C^{(diag)}_{\alpha}(p,q):=\left\langle \sum_{c=1}^M\frac{1}{|z-K_c|^2} \right\rangle=\left\langle \sum_{c=1}^M\frac{1}{(p-u_c)^2+(q+v_c)^2} \right\rangle,
 \end{equation}
and
\begin{equation}\label{corr2off}
 C^{(off)}_{\alpha}(p,q):=\left\langle \sum_{c\ne c'}^M\frac{1}{(z-K_c)(\overline{z}-\overline{K_{c'}}} \right\rangle
 \end{equation}
 \[
 =\left\langle \sum_{c\ne c'}^M\frac{1}{u_c-u_{c'}-i(2q+v_c+v_{c'})}\left(\frac{1}{p-u_c+i(q+v_c)}-\frac{1}{p-u_{c'}-i(q+v_{c'})}\right)\right\rangle
 \]
At the next step let us introduce the Fourier-transform in variable $p$:
\[
\tilde{C}_{\alpha}(k,q):=\frac{1}{2\pi}\int_{-\infty}^{\infty}e^{ipk} C_{\alpha}(p,q)\,dp.
\]
Taking into account $q>0, v_c\ge 0, \forall c$ we get
\begin{equation}\label{corr3diag}
 \tilde{C}^{(diag)}_{\alpha}(k,q)=\left\langle \frac{1}{2}\sum_{c=1}^M \frac{e^{iku_c-|k|(q+v_c)}}{q+v_c} \right\rangle
 \end{equation}
 whereas the Fourier transformed off-diagonal part reads as
 \begin{equation}\label{corr3off}
 \tilde{C}^{(off)}_{\alpha}(k,q)=\left\langle \sum_{c\ne c'}^M \frac{-i}{u_c-u_{c'}-i(2q+v_c+v_{c'})} \right.
 \end{equation}
 \[
 \left. \times \left(\theta(-k)e^{iku_c+k(q+v_c)}+\theta(k)e^{iku_{c'}-k(q+v_{c'})}\right) \right\rangle
 \]
where $\theta(k)=1$ for $k\ge 0$ and zero otherwise.

The next step is to continue analytically in the parameter $q$ from positive real values to the whole complex plane slit along the negative real line: $q=-v, v>0$, and
evaluate the jump across the slit, defined as

\begin{equation}\label{slit}
 \delta\tilde{C}_{\alpha}(k,v>0):=\lim_{\epsilon\to 0} \left(\tilde{C}_{\alpha}(k,-v-i\epsilon)-\tilde{C}_{\alpha}(k,-v+i\epsilon)\right)
 \end{equation}
For the diagonal part one finds after straightforward algebra
\begin{equation}\label{slitdiagA}
 \delta\tilde{C}^{(diag)}_{\alpha}(k,v>0)=i\lim_{\epsilon\to 0} \left\langle\sum_{c=1}^M e^{iku_c+|k|(v-v_c)}\left(\frac{\epsilon\cos(\epsilon |k|)}{\epsilon^2+(v-v_c)^2}-\frac{\sin(\epsilon |k|)(v-v_c)}{\epsilon^2+(v-v_c)^2}\right)\right\rangle
 \end{equation}
which upon using
\[
\lim_{\epsilon\to 0} \left(\frac{\epsilon \cos(\epsilon |k|)}{\epsilon^2+(v-v_c)^2}-\frac{\sin(\epsilon |k|)(v-v_c)}{\epsilon^2+(v-v_c)^2}\right)=\pi\delta(v-v_c)
\]
reduces the diagonal contribution to
\begin{equation}\label{slitdiagB}
 \delta\tilde{C}^{(diag)}_{\alpha}(k,v>0)=i\pi\left\langle\sum_{c=1}^M e^{iku_c}\,\delta(v-v_c)\right\rangle
 \end{equation}
At the same time straightforward computations show that assuming that the eigenvalues of the $K-$matrix are all distinct, i.e. $u_c-iv_c\ne u_c'-iv_{c'}$ for $c\ne c'$,
 the off-diagonal part does not generate any nonvanishing jump across the slit at $q=-v, v>0$, that is $\delta\tilde{C}^{(off)}_{\alpha}(k,v>0)=0$.
Finally, applying in Eq.(\ref{slitdiagB}) the inverse Fourier transform in the variable $k$ and comparing with the definition Eq.(\ref{denK})
provides the expression for the density of complex eigenvalues of the $K-$matrix in the form
\begin{equation}\label{slitdiagB}
\rho_M(u,v;y)= \frac{1}{2i\pi^2}\int_{-\infty}^{\infty}e^{-iku}\delta\tilde{C}_{\alpha=y\Delta/2\pi}(k,v>0)\,dk.
 \end{equation}
In this way the problem of computing  the density $\rho_M(u,v;y)$ is reduced to ability to evaluate explicitly the correlation function  $C_{\alpha}(p,q>0)$
in Eq.(\ref{corr1}) and perform the required Fourier transforms and jump evaluation. Below we show how this program is executed for those disordered or chaotic systems with broken time-reversal invariance which can be mapped onto the corresponding nonlinear $\sigma-$model.

\subsection{\small Computations for systems with broken time-reversal invariance}

Referring the interested reader to \cite{FSTresden} and references therein for a detailed discussion of physical assumptions behind such mapping, we just mention here that it provides the most powerful and systematic approaches to addressing universal single particle features of wave propagation in a disordered medium, including Anderson localization phenomena. Developed in the seminal works by Efetov \cite{Efe_book} building on earlier ideas of Wegner \cite{Wegner79} the model is defined by specifying a weight function
$e^{-{\cal S}[Q]}$, with the action ${\cal S}[Q]$ describing interaction between supermatrices $Q({\bf r})$  (i.e.\ matrices with Grassmann/anticommuting/ fermionic and ordinary/commuting/bosonic entries) associated
 to every site ${\bf r}\in \tilde{\mathbf{\Lambda}}$ located on an auxiliary lattice $\tilde{\mathbf{\Lambda}}$.  The size of supermatrices involved depends on the underlying symmetries of the Hamiltonian $H$, and in the simplest case of the Hamiltonians with fully broken time-reversal symmetry, denoted in the standard
nomenclature as class~A with Dyson parameter $\beta=2$, the supermatrices are of the size $4\times 4$. Physically such model provides, in a certain sense, a coarse-grained description of the original microscopic Anderson model or its continuous equivalent, with non-universal features on scales smaller than the mean-free path $l$
 being effectively integrated out. In such a picture every (super)matrix $Q({\bf r})$ associated to a single lattice site in $ \tilde{\mathbf{\Lambda}}$ "lumps together" behaviour of the microscopic model on scales of the order of the mean-free path $l$. From this point of view the billiards in the quantum chaotic regime, where essentially $l$ is of the same order as the system length $L$, are effectively characterized by nonlinear $\sigma-$models with a single matrix $Q$ without any spatial dependence.
 Such limit is traditionally called "zero-dimensional". At the same time
  all effects of the Anderson localization require considering extended lattices of interacting $Q-$matrices.

   One of the central objects of such theory turns out to be the
so called “order parameter function” (OPF) $F_{\bf r}(Q)$ which is formally defined \cite{Zirn86} by integrating the weight $e^{-{\cal S}[Q]}$ over all but one supermatrix
$Q({\bf r})$.
Due to global symmetries of the action, the OPF  can be shown to actually depend only on a few real Cartan variables parametrizing $Q$ matrices. In particular, for systems with broken time-reversal symmetry one has $F_{\bf r}(Q):={\cal F}(\lambda,\lambda_1)$, with $\lambda\in[-1,1]$ and $\lambda_1\in[1,\infty]$ being the compact and non-compact coordinates, respectively (we omitted spatial dependence on ${\bf r}$ for brevity).  Note that the OPF characterizes the {\it closed} system which (in the absence of absorption) conserves the number of particles, whereas allowing particles/waves at a given energy to be sent via the lead to the random medium and then collecting the reflected waves renders the medium {\it open}. However, if one makes an assumption of ``locality" of the lead, whose transverse extent is assumed to be much smaller
than the mean-free path $l$ in the disordered medium, makes the coupling to it effectively point-wise at the level of $\sigma$-model description. Still, even such point-wise lead may support arbitrary many propagation channels $M$, though we will be always assuming $M$ remaining negligible to the number of sites in the underlying microscopic lattice~$\mathbf{\Lambda}$.

The power of nonlinear $\sigma$-model description in our case lies in our ability to provide an explicit representation for the correlation function $ C_{\alpha}(p,q)$ defined in Eq.(\ref{corr1}) in terms of the OPF ${\cal F}(\lambda,\lambda_1)$ at the point of lead attachment. For systems with broken time-reversal invariance such computation has been already
performed in \cite{FyoSom97}, albeit formally only in the "zero-dimensional" limit, with OPF taking an especially simple form ${\cal F}(\lambda,\lambda_1)
=e^{-y(\lambda_1-\lambda)}$, where as before $y=2\pi \alpha/\Delta$ is the effective absorption parameter. It is however straightforward to adapt the calculation
for arbitrary nonlinear sigma-model, see Appendix B of \cite{Fyo05}, the result being given by the sum of two contributions, the disconnected one
\begin{equation}\label{corrSUSYdisc}
 C_{\alpha}^{(disc)}(p,q)={\small \sum_{c=1}^M\frac{1}{p-\gamma_c\frac{E}{2}-i(q+\pi\nu(E)\gamma_c)}\sum_{b=1}^M\frac{1}{p-\gamma_c\frac{E}{2}+i(q+\pi\nu(E)\gamma_b)}}
 \end{equation}
and the connected one
\begin{equation}\label{corrSUSYcon}
 C_{\alpha}^{(con)}(p,q)=\int_{-1}^1d\lambda\int_{1}^{\infty}d\lambda_1\frac{{\cal F}(\lambda,\lambda_1)}{(\lambda_1-\lambda)^2}\,{\cal R}_M(p,q|\lambda,\lambda_1)
 \end{equation}
where  the last factor in Eq.(\ref{corrSUSYcon}) is given by
\begin{equation}\label{corrG}
 {\cal R}_M(p,q|\lambda,\lambda_1):=\mathbf{\cal L}_{p,q}\prod_{c=1}^M\frac{(p-\gamma_c\frac{E}{2})^2+q^2+2\pi\nu(E)\gamma_c\lambda+(\pi\nu(E)\gamma_c)^2}
 {(p-\gamma_c\frac{E}{2})^2+q^2+2\pi\nu(E)\gamma_c\lambda_1+(\pi\nu(E)\gamma_c)^2}
 \end{equation}
with  the coupling coefficients $\gamma_c$  defined in Eq.(\ref{orth}) and  the differential operator $\mathbf{\cal L}_{p,q}:=\frac{1}{4}\left(\frac{\partial ^2}{\partial p^2}+\frac{\partial ^2}{\partial q^2}\right)$.

These expressions provide the basis for implementing the analytic continuation procedure described above.
For simplicity we consider below explicitly only the case $E=0$, so that $\pi \nu(E)=1$, and largely concentrate on the simplest, yet important case of equivalent channels: $\gamma_c=\gamma, \, \forall c=1,\ldots,M$ (see however Eq.(\ref{twochannonequiv}) for two non-equivalent channels). The analytic continuation procedure for the disconnected part amounts to a straightforward repetition of our derivation
of Eq.(\ref{slitdiagB}) and yields $\rho^{(disc)}(u,v)=\sum_{c=1}^M\delta(u)\delta(v-\gamma_c)$. The connected contribution to the density is much less trivial and we consider it below.

One starts with rewriting  Eq.(\ref{corrG}) in the form
\begin{equation}\label{corrGperf1}
 {\cal R}_M(p,q|\lambda,\lambda_1)=\mathbf{\cal L}_{p,q}\left(1-\frac{2q\gamma(\lambda_1-\lambda)}
 {p^2+q^2+2\gamma_c\lambda_1+\gamma_c^2}\right)^M,
 \end{equation}
 which after expanding the binomial reduces to
\begin{equation}\label{corrGperf2}
 {\cal R}_M(p,q|\lambda,\lambda_1)=-\sum_{l=1}^M\left(\begin{array}{c}M\\l\end{array}\right)\frac{(\lambda_1-\lambda)^l}{(l-1)!}
 \frac{\partial^{l-1}}{\partial \lambda_1^{l-1}}\mathbf{\cal L}_{p,q}\frac{2q\gamma}{p^2+q^2+2\gamma_c\lambda_1+\gamma_c^2}.
 \end{equation}
The latter form makes it an easy task to perform the Fourier transform in the variable $p$ assuming $q>0$, which essentially amounts to making in Eq.(\ref{corrGperf2}) the replacement
\[
\mathbf{\cal L}_{p,q}\frac{2q\gamma}{p^2+q^2+2\gamma_c\lambda_1+\gamma_c^2}\,\, \longrightarrow\,\, \phi(k,q)= \frac{\pi\gamma}{2}\left(\frac{\partial ^2}{\partial q^2} -k^2\right)\,q\frac{e^{-|k|\sqrt{q^2+2\gamma \lambda_1 q+\gamma^2}}}{\sqrt{q^2+2\gamma \lambda_1 q+\gamma^2}}
\]
Following the procedures described in Eq(\ref{slit}) we now continue analytically in the parameter $q$ from positive real values to the whole complex plane slit along the negative real line: $q=-v, v>0$, and
evaluate the associated jump across the slit
\begin{equation}\label{slitphiA}
 \delta\phi(k,v>0):=\lim_{\epsilon\to 0} \left(\phi(k,-v-i\epsilon)-\phi(k,-v+i\epsilon)\right)
 \end{equation}
 which is easily found to be equal to
\begin{equation}\label{slitphiB}
{\small
\delta\phi(k,v>0)=\pi\gamma\left(\frac{\partial ^2}{\partial v^2}-k^2\right) v \frac{\cos{k\sqrt{2\gamma \lambda_1 v-v^2-\gamma^2}}}{\sqrt{2\gamma \lambda_1 v-v^2-\gamma^2}}\,\theta(2\gamma \lambda_1 v-v^2-\gamma^2).
}
\end{equation}
Straightforward inversion of the Fourier-transform in the variable $k$ converts the above into
\[
\delta\phi(u,v>0)=\frac{\gamma}{2}\left(\frac{\partial ^2}{\partial u^2}+\frac{\partial ^2}{\partial v^2}\right) v
\]
\begin{equation*}
\times
\frac{\delta\left(u-\sqrt{2\gamma \lambda_1 v-v^2-\gamma^2}\right)+\delta\left(u+\sqrt{2\gamma \lambda_1 v-v^2-\gamma^2}\right)}{\sqrt{2\gamma \lambda_1 v-v^2-\gamma^2}}\,\theta(2\gamma \lambda_1 v-v^2-\gamma^2).
\end{equation*}
\begin{equation}\label{slitphiB}
=\frac{1}{2}\left(\frac{\partial ^2}{\partial u^2}+\frac{\partial ^2}{\partial v^2}\right) \delta\left(\lambda_1-x_{\gamma}\right), \quad x_{\gamma}:=\frac{u^2+v^2+\gamma^2}{2\gamma\,v}.
\end{equation}
Next we trade the derivatives over $\lambda_1$ for those over $x_{\gamma}$ by the identity
\[
 \frac{\partial^{l-1}}{\partial \lambda_1^{l-1}}\delta\left(\lambda_1-x_{\gamma}\right)=(-1)^{l-1} \frac{\partial^{l-1}}{\partial x_{\gamma}^{l-1}}\delta\left(\lambda_1-x_{\gamma}\right)
\]
and in this way arrive at replacing Eq.(\ref{corrGperf2}) with
\begin{equation}\label{corrGperf22}
 \tilde{{\cal R}}_M(u,v|\lambda,\lambda_1)=-\frac{1}{2}\left(\frac{\partial ^2}{\partial u^2}+\frac{\partial ^2}{\partial v^2}\right)\sum_{l=1}^M\left(\begin{array}{c}M\\l\end{array}\right)\frac{(-1)^{l-1}(\lambda_1-\lambda)^l}{(l-1)!}
 \frac{\partial^{l-1}}{\partial x_{\gamma}^{l-1}}\delta\left(\lambda_1-x_{\gamma}\right).
 \end{equation}
With this Eqs. (\ref{slitdiagB}) and Eq.(\ref{corrSUSYcon}) imply the density of $K-$matrix eigenvalues via
\begin{equation}\label{denKcontA}
\rho^{(con)}(u,v)=\frac{1}{2\pi}\int_{-1}^1d\lambda\int_{1}^{\infty}d\lambda_1\frac{{\cal F}(\lambda,\lambda_1)}{(\lambda_1-\lambda)^2}\,\tilde{{\cal R}}_M(u,v|\lambda,\lambda_1)
\end{equation}
which upon substituting (\ref{corrGperf22}) into it and changing the order of integrations yields
\begin{equation}\label{denKcontA}
\rho^{(con)}(u,v)=\frac{1}{4\pi}\left(\frac{\partial ^2}{\partial u^2}+\frac{\partial ^2}{\partial v^2}\right)\int_{-1}^1d\lambda\, G_M(\lambda|x_{\gamma}).
\end{equation}
Here we denoted
\begin{equation}\label{denKcontB}
 G_M(\lambda|x_{\gamma}):=\sum_{l=1}^M\left(\begin{array}{c}M\\l\end{array}\right)\frac{(-1)^{l-1}}{(l-1)!}
 \frac{\partial^{l-1}}{\partial x_{\gamma}^{l-1}}\left[(x_{\gamma}-\lambda)^lT(\lambda|x_{\gamma})\right],
\end{equation}
with
\begin{equation}\label{denKcontBT}
T(\lambda|x_{\gamma})=\frac{{\cal F}(\lambda,x_{\gamma})}{(x_{\gamma}-\lambda)^2}.
\end{equation}
Applying the Leibnitz formula
\[
 \frac{\partial^{l-1}}{\partial x_{\gamma}^{l-1}}\left[(x_{\gamma}-\lambda)^lT(\lambda|x_{\gamma})\right]=\sum_{k=0}^{l-1}\left(\begin{array}{c}l-1\\k\end{array}\right)
 \frac{l!}{(k+1)!}(x_{\gamma}-\lambda)^{k+1}\frac{\partial^k}{\partial x_{\gamma}^k}T(\lambda|x_{\gamma})
\]
and substituting it back to Eq.(\ref{denKcontB}) one may change the order of summation as
\[
 G_M(\lambda|x_{\gamma})=\sum_{l=1}^MA_l\sum_{k=0}^{l-1}B_{k,l}V_k=\sum_{k=0}^{M-1}V_k\sum_{l=k+1}^MA_lB_{k,l}
\]
with $ V_k:=(x_{\gamma}-\lambda)^{k+1}\frac{\partial^k}{\partial x_{\gamma}^k}T(\lambda|x_{\gamma})$ and
\[
A_l:=\left(\begin{array}{c}M\\l\end{array}\right)\frac{(-1)^{l-1}}{(l-1)!}, \quad B_{k,l}:=\left(\begin{array}{c}l-1\\k\end{array}\right)
 \frac{l!}{(k+1)!}.
\]
This gives
\[
\sum_{l=k+1}^MA_lB_{k,l}=\frac{M!}{k!(k+1)!}\sum_{l=k+1}^M\frac{(-1)^{l-1}}{(M-l)!}\frac{1}{(l-k-1)!}
\]
\[=\frac{(-1)^k\,M!}{(M-k-1)!k!(k+1)!}
\sum_{n=0}^{M-k-1}(-1)^n\left(\begin{array}{c}M-k-1\\n\end{array}\right)=\frac{(-1)^{M-1}}{(M-1)!}\delta_{k,M-1}
\]
using the Kronecker symbol $\delta_{k,k'}$,  since the sum over $n$ is vanishing for all $0\le k<M-1$, and is equal to unity at $k=M-1$.

As the result we get the final expression for the connected part of the mean density of $K-$matrix eigenvalues in the form
\begin{equation}\label{denKcontFINA}
\rho^{(con)}_M(u,v)=\frac{1}{4\pi}\frac{(-1)^{M-1}}{(M-1)!}\left(\frac{\partial ^2}{\partial u^2}+\frac{\partial ^2}{\partial v^2}\right)\int_{-1}^1d\lambda\,(x_{\gamma}-\lambda)^M
\, \frac{\partial^{M-1}}{\partial x_{\gamma}^{M-1}}\frac{{\cal F}(\lambda,x_{\gamma})}{(x_{\gamma}-\lambda)^2}
\end{equation}

A few remarks are here in order which help to properly interpret and appreciate the content of Eq.(\ref{denKcontFINA}). \\
{\bf Remark 1.} Recalling from Eq.(\ref{slitphiB}) that
\begin{equation}\label{xdef}
x_{\gamma}=\frac{u^2+v^2+\gamma^2}{2\gamma\,v}\equiv \frac{u^2}{2\gamma\,v}+\frac{1}{2}\left(\frac{v}{\gamma}+\frac{\gamma}{v}\right)\ge 1
\end{equation}
one may straightforwardly check that for any smooth enough function $\Phi(x)$ holds
\begin{equation}\label{identdiff}
\left(\frac{\partial ^2}{\partial u^2}+\frac{\partial ^2}{\partial v^2}\right)\Phi(x_{\gamma})=\frac{1}{v^2}\frac{\partial}{\partial x_{\gamma}}(x_{\gamma}^2-1)\frac{\partial}{\partial x_{\gamma}}\Phi(x_{\gamma}), \quad x_{\gamma}>1.
\end{equation}
This was exactly the form used to represent the density in \cite{FSTresden}.\\
 There is however a subtlety in Eq.(\ref{denKcontFINA}) related with its content at $x_{\gamma}\to 1$.
In our derivation we tacitly assumed $x_{\gamma}>1$. However, a more careful analysis shows that the integral
in the right-hand side of  Eq.(\ref{denKcontFINA}) should be pre-multiplied with the step-function factor $\theta(x_{\gamma}-1)$
 arising as the result of performing integration over $\lambda_1\in [1,\infty)$ with the factor $\delta(\lambda_1-x_{\gamma})$.   Presence of such a seemingly innocent $\theta-$factor has however important consequences: when acted upon with the differential operator in the right-hand side of Eq.(\ref{identdiff}) it generates
 the $\delta$-function factors exactly cancelling the contribution from the disconnected part, $\rho^{(disc)}(u,v)=\sum_{c=1}^M\delta(u)\delta(v-\gamma_c)$.
 As a result, the formula  Eq.(\ref{denKcontFINA}) as it is written (i.e. without $\theta-$factor)  in fact gives the full, properly normalized,  eigenvalue density for
 the $K-$matrix in absorptive systems. A similar mechanism of cancellation of $\delta-$terms has been first noticed in \cite{rozhkov2003statistics}, and we explain  in the Appendix A how it works in our case using the simplest case of $M=1$ as an example.\\

{\bf Remark 2.}  With the hindsight, one may notice that one could have arrived to the same expression Eq.(\ref{denKcontFINA}) by a much simpler procedure.
Namely, by defining
\begin{equation}\label{xtilddef}
\tilde{x}:=\frac{p^2+q^2+\gamma^2}{2\gamma\,q}
\end{equation}
rewrite Eq.(\ref{corrGperf1}) in the form
\begin{equation}\label{corrGperf1xtild}
 {\cal R}_M(p,q|\lambda,\lambda_1)=\frac{1}{4q^2}\frac{\partial}{\partial\tilde{x}}(\tilde{x}^2-1)
 \frac{\partial}{\partial\tilde{x}}\left(\frac{\tilde{x}+\lambda} {\tilde{x}+\lambda_1}\right)^M,
 \end{equation}
Then simply replace $u\to p, q\to -v-i0$ implying $\tilde{x}\to -x_{\gamma}+i0$ and calculate the associated jump across the cut using
\[
\Im \left(\frac{-x_{\gamma}+i0+\lambda} {-x_{\gamma}+i0+\lambda_1}\right)^M=(x_{\gamma}-\lambda)^M\frac{(-1)^{M-1}}{(M-1)!}\frac{\partial^{M-1}}{\partial x_{\gamma}^{M-1}}
\Im \frac{1}{x_{\gamma}-\lambda_1-i0}
\]
\[
=\pi(x_{\gamma}-\lambda)^M\frac{(-1)^{M-1}}{(M-1)!}\frac{\partial^{M-1}}{\partial x_{\gamma}^{M-1}}\delta(x_{\gamma}-\lambda_1)
\]
Such recipe was exactly one employed for $M=1$ in \cite{SavSomFyo2005}, though without a proper explanation provided there or in the review \cite{Fyo05}.

Armed with such a recipe, one can easily apply it to the case of non-equivalent channels. General formulas look in that case quite complicated, but in
the simplest case of two non-equivalent channels with coupling constants $\gamma_1\ne \gamma_2$ one gets a relatively compact expression:
\begin{equation}\label{twochannonequiv}
\rho_{\gamma_1,\gamma_2}(u,v)=\frac{1}{4\pi}\left(\frac{\partial ^2}{\partial u^2}+\frac{\partial ^2}{\partial v^2}\right)\left\{-\int_{-1}^1\,d\lambda\,
\frac{{\cal F}(\lambda,x_{1})-{\cal F}(\lambda,x_{2})}{x_{1}-x_2}\right.
\end{equation}
\[
\left.+\int_{-1}^1\,d\lambda \left[\frac{{\cal F}(\lambda,x_{1})}{x_2-\lambda_2}+\frac{{\cal F}(\lambda,x_{2})}{x_1-\lambda_1}\right]\right\},
\]
where we defined
\begin{equation}\label{xdefnoneq}
x_{1}=\frac{u^2+v^2+\gamma_{1}^2}{2\gamma_1\,v}, \quad x_{2}=\frac{u^2+v^2+\gamma_{2}^2}{2\gamma_2\,v}.
\end{equation}
{\bf Remark 3.}  It is clear that performing further analysis of Eq.(\ref{denKcontFINA}) hinges on our ability to have a good understanding of
the OPF ${\cal F}(\lambda,x)$ for the closed counterpart of the scattering system, which in general also depends on the (appropriately normalized) absorption parameter $\alpha$. Such knowledge is currently available mainly in two cases (i) the "zero-dimensional" limit, with OPF taking an especially simple form ${\cal F}^{(0d)}(\lambda,x)
=e^{-y(x-\lambda)}$, where as before $y=2\pi \alpha/\Delta$ and (ii) in a (semi) infinite quasi-one dimensional wire, see the sketch below, of length $L\to \infty$,
with one edge closed for the waves and second edge attached to an infinite waveguide with $M$ propagating channels.
\begin{figure}
\centering
\includegraphics[width=100mm]{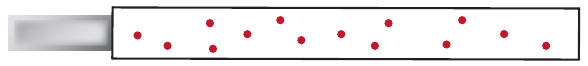}
\caption{\small A sketch of the “quasi-1D” model. The left part in grey represents an infinite-length ideal lead supporting $M$
propagating modes. The disordered part is of a finite length $L$ and contains finite concentration of random impurities inside.}
\label{F:quasi1d}
\end{figure}
Such wire is characterized by a classical microscopic diffusion constant
$D$ related to the localization length $\xi$ of quantum wave problem as $\xi=2\pi\nu D$, with $\nu$ being as before the mean eigenvalue density at a given energy.
Note that mathematically such wires can be modelled by a large banded random matrix \cite{FyoMirRBM91,FyoMirRBM94}.
In such a system the OPF at points close to its edges has been originally found in \cite{SO07} and takes the following form in terms of the modified Bessel functions $I_p(z),K_p(z)$:
\be \label{zeromode}
  {\cal F}^{(1d)}(\lambda,x)
  =
  K_0(a) b I_1(b)+a K_1(a) I_0(b) ,
\ee
with
\be \label{pq_def}
  a = \kappa \sqrt{(x +1)/2} ,
\qquad
  b = \kappa \sqrt{(\lambda +1)/2} ,
\ee
where the parameter $\kappa$ is related to the absorption $\alpha$ as
\be
\label{kappa-def}
  \kappa = \sqrt{8\alpha/\Delta_\xi},
\ee
with an important energy scale $\Delta_\xi = (4\pi^2D\nu^2)^{-1}=D/\xi^2$ giving the mean level spacing in the quasi-one dimensional wires whose length $L$ equal to the localization length $\xi$.

In the "zero-dimensional" limit, due to a simple form of the OPF one can relatively straightforwardly perform the required integrations and differentiations
in  Eq.(\ref{denKcontFINA}) and get the explicit formulas, which we present below for the simplest cases $M=1$ and $M=2$ of equivalent channels:
  \begin{equation}\label{singlechan0d}
\rho_{0d,M=1}(u,v)=\frac{1}{2\pi v^2}e^{-x_{\gamma}}\left(y\cosh{y}-\sinh{y}(1-yx_{\gamma})\right)
\end{equation}
and
\begin{equation}\label{2chan0dequiv}
\rho_{0d,M=2}(u,v)=\frac{1}{2\pi v^2}e^{-x_{\gamma}}\left(\sinh{y}[y(x^2_{\gamma}-1)-2x_{\gamma}]+2[y\cosh{y}-\sinh(1-yx_{\gamma})]\right).
\end{equation}
with the same definition of $x_{\gamma}$, Eq.(\ref{xdef}).
The formula equivalent to Eq.(\ref{singlechan0d}) appeared already in the literature, see Eq.(5) in \cite{FyoSav2004}, the two channel case seems to be new.
As to the quasi-$1D$ system of infinite length, it turns out that again the results can be found explicitly in the general case.
 Below we present it only for the simplest case of a single attached channel, when the density aquires quite an elegant form after manipulations outlined in the Appendix B to this paper:
\begin{equation}\label{singlechan1d}
\rho_{1d,M=1}(u,v)=\frac{1}{2\pi v^2}P_0(x_{\gamma}), \quad P_0(x)=\frac{\kappa^2}{4}\left[I_2(\kappa)K_0\left(\kappa\sqrt{\frac{x+1}{2}}\right)\right.
\end{equation}
\[
\left.+I_1(\kappa)\sqrt{\frac{x+1}{2}}K_1\left(\kappa\sqrt{\frac{x+1}{2}}\right)\right]
\]
As is shown in \cite{FyoSav2004}, for $M=1$ and $\gamma=1$ the variable $r=(x-1)/(x+1)$ is nothing else but the modulus of the reflection coefficient, which in the absorptive system is smaller than one. Correspondingly, the function $P_0(x)$ in Eq.(\ref{singlechan1d}) provides the distribution for $x$, hence for $r$, in a single-channel quasi-$1D$
system with absorption. This complements a  result for the same geometry in the case of no absorption inside the sample, but for the second edge of the sample being in contact with perfectly absorbing lead, see eqs. (12)-(13) in \cite{Fyo2003}. Note also that it is not difficult  to integrate further out the variable $u$, getting an explicit formula for the distribution of variable $v$, known as the local density of states, corresponding to locations close to the sample edge. The latter is an important characteristic of disordered single-particle systems, see \cite{MirFyo1994B,MirFyo1994A,LDOS_num_2010}.

In conclusion, we derived the mean density of complex eigenvalues for random Wigner reaction $K-$matrices for absorptive
disordered or chaotic systems with broken time-reversal invariance,  in the sigma-model approximation.
Extension of these results to systems with preserved time-reversal invariance (and then eventually symplectic symmetry) is certainly possible along similar lines, generalizing
$M=1$ results presented in \cite{Fyo05}. These subjects are left for future publications.

{\bf Acknowledgments:}
This research has been supported by the EPSRC Grant EP/V002473/1 ``Random Hessians and Jacobians: theory and applications''.

\vspace{1cm}

\appendix{\bf Appendix A: cancellation of the disconnected part.}\\
Our starting point  is the formula Eq.(\ref{denKcontFINA}) with included $\theta-$factor,  specified for simplicity and transparency to
 the case of a single channel $M=1$ and $\gamma=1$, so that $x_{\gamma=1}=x$. We write it in the form
\begin{equation}\label{denKcontFINA_M=1}
\rho^{(con)}_{M=1}(u,v)=\frac{1}{4\pi v^2}{\cal L}_{x} \left[\theta(x-1) \Phi(x)\right], \quad \Phi(x)=\int_{-1}^1\,\frac{{\cal F}(\lambda,x)}{(x-\lambda)}\,d\lambda,
\end{equation}
where we introduced the differential operator
\be\label{Legendre_dif}
{\cal L}_{x} :=\frac{\partial}{\partial x}(x^2-1)\frac{\partial}{\partial x}.
\ee
Straightforward differentiation then gives
\begin{equation}\label{difftheta1}
{\cal L}_{x} \left[\theta(x-1) \Phi(x)\right]=\delta(x-1)\left[2x\Phi(x)+2(x^2-1)\Phi'(x)\right]
\end{equation}
\[
+\delta'(x-1)\left[(x^2-1)\Phi(x)\right]+\theta(x-1){\cal L}_{x} \Phi(x).
\]
Further using the integration by parts identity
{\small
\[
\delta'(x-1)\left[(x^2-1)\Phi(x)\right]=-\delta(x-1)\frac{d}{dx}\left[(x^2-1)\Phi(x)\right]=-\delta(x-1)\left[2x\Phi(x)+(x^2-1)\Phi'(x)\right]
\]
}
we conclude that
\begin{equation}\label{difftheta1}
{\cal L}_{x} \left[\theta(x-1) \Phi(x)\right]=\delta(x-1)\left[(x^2-1)\Phi'(x)\right]+\theta(x-1){\cal L}_{x} \Phi(x)
\end{equation}
so it remains to evaluate $\lim_{x\to 1}\left[(x^2-1)\Phi'(x)\right]$. To this end we notice that it can be generally shown that
 $\lim_{x\to 1}{\cal F}(\lambda,x)=1$, hence from Eq.(\ref{denKcontFINA_M=1}) we have $\Phi(x\to 1)\approx \int_{-1}^1\,\frac{1}{(x-\lambda)}\,d\lambda
 =\ln{\frac{x+1}{x-1}}$, which immediately implies $\lim_{x\to 1}\left[(x^2-1)\Phi'(x)\right]=-2$. This gives the singular contribution to the density Eq.(\ref{denKcontFINA_M=1}) in terms of the variables $u,v$ given by
\[
-\frac{2}{4\pi v^2}\delta\left(\frac{u^2+v^2+1}{2v}-1\right)=-\frac{1}{\pi v}\delta\left(u^2+(v-1)^2\right)=-\delta(u)\delta(v-1)
\]
which exactly cancels the contribution from the disconnected part.

\vspace{1cm}
\appendix{\bf Appendix B}\\
In this appendix we show how Eq.(\ref{zeromode}) when substituted to Eq.(\ref{denKcontFINA}) implies Eq.(\ref{singlechan1d}).
Throughout this appendix we again use $x_{\gamma}=x$ and ${\cal L}_x:=\frac{\partial}{\partial x}(x^2-1)\frac{\partial}{\partial x}$.
First of all, we use the identity (43) from the paper \cite{FSTresden}, which claims that
\be\label{difident1}
\frac{\partial}{\partial \kappa} {\cal F}^{(1d)}(\lambda,x)=-\frac{\kappa}{2}(x-\lambda)K_0\left(\kappa\sqrt{\frac{x+1}{2}}\right)
I_0\left(\kappa\sqrt{\frac{\lambda+1}{2}}\right)
\ee
By differentiating both sides of Eq.(\ref{denKcontFINA}) over $\kappa$ and using Eq.(\ref{difident1}) in the right-hand side yields
\be\label{deri1}
\frac{\partial}{\partial \kappa}\rho_{1d,M=1}(u,v)=-\frac{1}{8\pi v^2}{\cal L}_x\left[\kappa\,K_0\left(\kappa\sqrt{\frac{x+1}{2}}\right)\int_{-1}^1
I_0\left(\kappa\sqrt{\frac{\lambda+1}{2}}\right)\,d\lambda\right]
\ee
and after performing the integral by substitution $\lambda=2z^2-1, z\in[0,1]$ find that
\be\label{deri2}
\frac{\partial}{\partial \kappa}\rho^{(con)}_{1d,M=1}(u,v)=-\frac{1}{2\pi v^2}{\cal L}_x\left[K_0\left(\kappa\sqrt{\frac{x+1}{2}}\right)
I_1\left(\kappa\right)\right]
\ee
\be\label{deri3}
=-\frac{1}{2\pi v^2}\frac{\partial}{\partial x}\sqrt{\frac{x+1}{2}}\left[\frac{1-x}{2}\,\kappa I_1\left(\kappa\right)\,K_1\left(\kappa\sqrt{\frac{x+1}{2}}\right)
\right]
\ee
At the next step we employ the following identity (c.f.  5.54 in p.624 of \cite{GRbook}):
\be\label{GRident}
\frac{1-x}{2}\,\kappa I_1\left(\kappa\right)\,K_1\left(\kappa\sqrt{\frac{x+1}{2}}\right)=\frac{\partial}{\partial \kappa}\left[
\kappa I_2\left(\kappa\right)K_1\left(\kappa\sqrt{\frac{x+1}{2}}\right)\right.
\ee
\[
\left. + \sqrt{\frac{x+1}{2}}\kappa I_1\left(\kappa\right)K_2\left(\kappa\sqrt{\frac{x+1}{2}}\right)\right]
\]
Using the fact that $\rho^{(con)}_{1d,M=1}(u,v)\to 0 $ as $\kappa\to \infty$ we then may conclude that Eq.(\ref{deri3}) and Eq.(\ref{GRident}) together imply
{\small
\[
\rho_{1d,M=1}(u,v)=-\frac{1}{2\pi v^2}\frac{\partial}{\partial x}\sqrt{\frac{x+1}{2}}
\left\{\kappa I_2\left(\kappa\right)K_1\left(\kappa\sqrt{\frac{x+1}{2}}\right)+\sqrt{\frac{x+1}{2}}\kappa I_1\left(\kappa\right)K_2\left(\kappa\sqrt{\frac{x+1}{2}}\right)\right\}
\]
\[
=-\frac{1}{2\pi v^2}\left\{\frac{I_1\left(\kappa\right)}{\kappa}\frac{\partial}{\partial x}\left[\kappa^2\frac{x+1}{2}K_2\left(\kappa\sqrt{\frac{x+1}{2}}\right)\right]
+I_2\left(\kappa\right)\frac{\partial}{\partial x}\left[\kappa\sqrt{\frac{x+1}{2}}K_1\left(\kappa\sqrt{\frac{x+1}{2}}\right)\right]\right\}
\]
}
Finally introducing in the above the variable $z=\kappa\sqrt{\frac{x+1}{2}}$, using the chain rule and the identity (see 8.846.14 in \cite{GRbook})
\[
\frac{d}{dz}\left(z^pK_p(z)\right)=-z^{p}K_{p-1}(z)
\]
allows to bring the density $\rho^{(con)}_{1d,M=1}(u,v)$ to the final form Eq.(\ref{singlechan1d}).

\end{document}